\documentclass[11pt,amsfonts,fullpage]{amsart}

\usepackage{fullpage}

\def\endproof{\qed \medskip}
\def\blacksquare{\hbox to .60em{\vrule width .60em height .60em}}

\newtheorem{theorem}{Theorem}[section]

\newtheorem{proposition}[theorem]{Proposition}

\newtheorem{remark}[theorem]{Remark}

\begin{document}

\title[]{Existence and stability of even dimensional asymptotically de Sitter spaces}

\author[]{Michael T. Anderson}

\thanks{Partially supported by NSF Grant DMS 0305865}

\abstract{A new proof of Friedrich's theorem on the existence and stability 
of asymptotically de Sitter spaces in 3+1 dimensions is given, which extends 
to all even dimensions. In addition we characterize the possible limits of 
spaces which are globally asymptotically de Sitter, to the past and future.} 
\endabstract

\maketitle

\setcounter{section}{0}

\section{Introduction.}
\setcounter{equation}{0}

 Consider globally hyperbolic vacuum solutions $(M^{n+1}, g)$ to the 
Einstein equations with cosmological constant $\Lambda > 0$, so that 
\begin{equation} \label{e1.1}
Ric_{g} - \frac{R_{g}}{2}g + \Lambda g = 0. 
\end{equation}
The simplest solution is (pure) de Sitter space on $M^{n+1} = {\mathbb R} 
\times S^{n}$, with metric 
\begin{equation} \label{e1.2}
g_{dS} = -dt^{2} + \cosh^{2}(t) g_{S^{n}(1)}. 
\end{equation}
More generally, let $(N^{n}, g_{N})$ be any compact Riemannian manifold with 
metric $g_{N}$ satisfying the Einstein equation $Ric_{g_{N}} = (n-1)g_{N}$. 
Then the (generalized) de Sitter metric 
\begin{equation} \label{e1.3}
g_{dS}^{N} = -dt^{2} + \cosh^{2}(t) g_{N},
\end{equation}
on ${\mathbb R}\times N$ is also a solution of (1.1), with $\Lambda = n(n-1)/2$.

 Let $dS^{+}$ be the space of all globally hyperbolic spacetimes $(M^{n+1}, g)$ 
satisfying (1.1), with a spatially compact Cauchy surface, which are asymptotically 
de Sitter (dS) to the future, i.e. future conformally compact in the sense of Penrose; 
the terminology asymptotically simple is also used in this context. Thus there is a 
smooth function $\Omega$ such that the conformally compactified metric 
\begin{equation} \label{e1.4}
\bar g = \Omega^{2}g, 
\end{equation}
extends to the compactified spacetime $\bar M = M \cup {\mathcal I}^{+}$, 
where ${\mathcal I}^{+}$ is a compact $n$-manifold without boundary. 
The function $\Omega$ is smooth on $\bar M$, with $\Omega > 0$, 
${\mathcal I}^{+} = \Omega^{-1}(0)$ and $d\Omega  \neq 0$ on ${\mathcal I}^{+}$. 
The boundary metric $\gamma  = \bar g|_{{\mathcal I}^{+}}$ 
depends on the choice of $\Omega$; however the conformal class 
$[\gamma]$ of $\gamma$ is independent of $\Omega$ and is called 
future conformal infinity. Such spacetimes are geodesically complete 
to the future of an initial compact Cauchy surface $\Sigma$ diffeomorphic to 
${\mathcal I}^{+}$. There are no restrictions on the class $[\gamma]$ 
or the topology of ${\mathcal I}^{+}$, and so such spacetimes are sometimes 
also called asymptotically locally de Sitter. Changing the time orientation gives 
the same notion for $dS^{-}$, with future conformal infinity ${\mathcal I}^{+}$ replaced 
by past conformal infinity ${\mathcal I}^{-}$. The smoothness of $\bar g$ 
in (1.4) up to $\bar M$ may be measured in H\"older spaces $C^{m,\alpha}$, 
but we will mostly use Sobolev spaces $H^{s}$ which are more natural in this 
context. 

 In addition, let $dS^{\pm}$ be the space of such globally hyperbolic 
spacetimes which are in both $dS^{+}$ and $dS^{-}$; thus such spacetimes are 
(completely) global, in the sense that they are geodesically complete 
and asymptotically simple both to the past and to the future. 

\medskip

 Mathematically, the most significant result on the structure of such 
spacetimes is Friedrich's theorem [8], [9] that in $3+1$ dimensions, the 
Cauchy problem with data on ${\mathcal I}^{+}$, (or ${\mathcal I}^{-})$ is 
well-posed, cf. also [10] for recent discussions. Thus, for arbitrary 
Cauchy data on ${\mathcal I}^{+}$, there is a unique spacetime $(M^{4}, g)$ 
which realizes this data at future infinity. Moreover, small but 
arbitrary variations of the Cauchy data give rise to small 
perturbations of the solution. It follows in particular that the space 
$dS^{\pm}$ of global solutions is open; thus spaces in $dS^{\pm}$, in 
particular pure de Sitter space $(M^{4}, g_{dS})$, are stable under 
small perturbations of the Cauchy data at ${\mathcal I}^{+}$, (or 
${\mathcal I}^{-}$). The same statement holds for perturbations of the 
data on a compact Cauchy surface $\Sigma$ for $(M^{4}, g)$. 

 The purpose of this paper is to extend Friedrich's theorem to 
arbitrary even dimensions. Let ${\mathcal I}^{+}$ be any closed $n$-manifold, 
$n$ odd, and let $\gamma$ be any $H^{s+n}$ smooth Riemannian metric on 
${\mathcal I}^{+}, s > \frac{n}{2}+1$. Next, let $\tau$ be any $H^{s}$ 
symmetric bilinear form on ${\mathcal I}^{+}$ satisfying the constraints
\begin{equation} \label{e1.5}
tr_{\gamma}\tau  = 0, \ \ \delta_{\gamma}\tau  = 0, 
\end{equation}
i.e. $\tau$ is transverse-traceless with respect to $\gamma$. Define 
$\gamma_{1} \sim \gamma_{2}$ and $\tau_{1} \sim \tau_{2}$ if these data 
are conformally related, i.e. there exists $\lambda : {\mathcal I}^{+} 
\rightarrow  {\mathbb R}^{+}$ such that $\gamma_{2} = \lambda^{2}\gamma_{1}$ 
and $\tau_{2} = f(\lambda)\tau_{1}$, where $f$ is chosen so that (1.5) holds 
for $\tau_{2}$, cf. [5] for the exact transformation formula. Let 
$([\gamma], [\tau])$ be the equivalence class of $(\gamma , \tau)$. 
Then Cauchy data for the Einstein equations (1.1) with $\Lambda  > 0$ 
consist of triples $({\mathcal I}^{+}, [\gamma], [\tau])$. The form $\tau$ 
corresponds to the order $n$ behavior of the metric; roughly for $\bar g$ 
as in (1.3), $\tau  = (\partial_{\Omega})^{n}\bar g|_{{\mathcal I}^{+}}$; 
see \S 2 for further details.
\begin{theorem} \label{t 1.1.}
  The Cauchy problem for the Einstein equations with Cauchy data 
$({\mathcal I}^{+}, [\gamma], [\tau])$ at future conformal infinity is 
well-posed in $H^{s+n}\times H^{s}$, for any $s > \frac{n}{2}+2$.
\end{theorem}

 Thus, given any Cauchy data $([\gamma], [\tau]) \in  H^{s+n}({\mathcal 
I}^{+})\times H^{s}({\mathcal I}^{+})$ satisfying (1.5), up to isometry 
there is a unique Einstein metric $(M^{n+1}, g) \in  dS^{+}$ whose conformal 
compactification as in (1.4) induces the given data $([\gamma], [\tau])$ 
on ${\mathcal I}^{+}$.

\medskip

 This result has the following simple consequence:
\begin{theorem} \label{t 1.2.}
  The space $dS^{\pm}$ is open with respect to the $H^{s+n}\times H^{s}$ 
topology on ${\mathcal I}^{+}$, $s > \frac{n}{2}+2$. Thus, given any dS 
solution $(M^{n+1}, g_{0}) \in  dS^{\pm}$, any $H^{s+n}\times H^{s}$ small 
perturbation of the Cauchy data $([\gamma], [\tau])$ on ${\mathcal I}^{+}$ 
(or ${\mathcal I}^{-})$ gives rise to complete solution $(M^{n+1}, g) \in  dS^{\pm}$ 
globally close to $(M^{n+1}, g_{0})$. In particular, 
the even-dimensional pure de Sitter spaces $g_{dS}^{N}$ in (1.3) are globally stable. 
\end{theorem}

 Here, globally close is taken with respect to a natural $H^{s}$ topology 
on the conformal compactification $\bar M = M \cup  {\mathcal I}^{+} \cup  
{\mathcal I}^{-}$, see the proof for details. The complete solution $(M^{n+1}, g)$ 
induces $H^{s+n}\times H^{s}$ Cauchy data at both past and future conformal 
infinity ${\mathcal I}^{-}$, ${\mathcal I}^{+}$. Of course the size of the 
allowable perturbations in Theorem 1.2 depends on $(M^{n+1}, g_{0})$. 

\medskip

 We describe briefly the main ideas in the proof of Theorem 1.1; full 
details are given in \S 2. The Einstein equations (1.1) induce a $2^{\rm nd}$ 
order system of equations for a compactified metric $\bar g$ in 
(1.4). However, this system is degenerate at ${\mathcal I}^{+} = \{\Omega  
= 0\}$ and this degeneracy causes severe problems in trying to prove 
the well-posedness of the system. In $3+1$ dimensions, Friedrich [8] 
has developed a larger and more complicated system of evolution 
equations, the conformal Einstein equations, for the (unphysical) 
metric $\bar g$ together with other variables. This expanded 
system is non-degenerate and shown to be symmetric hyperbolic; then 
standard results on such systems lead to the well-posedness of the 
conformal field equations. However, it seems very unlikely that this 
method could succeed in higher dimensions, cf. [10], due at least in 
part to the special form of the Bianchi equations in $3+1$ dimensions. 

 The approach taken here is to replace the Einstein equation by a more 
complicated but conformally invariant higher order equation for the 
metric alone, whose solutions include the vacuum Einstein metrics (with 
$\Lambda$ term). In $3+1$ dimensions, this system is the system of 
$4^{\rm th}$ order Bach equations, cf. (2.12) below. The Bach equations 
have been used in a number of contexts in connection with issues related 
to conformal infinity, cf. [14], [15], [16], for example. 

 In higher even dimensions, in place of the Bach tensor, we use the 
ambient obstruction tensor ${\mathcal H}$ of Fefferman-Graham [6], which 
agrees with the Bach tensor in $3+1$ dimensions; this tensor is also 
characterized as the stress-energy tensor of the conformal anomaly, cf. 
[5]. The tensor ${\mathcal H}$ is a symmetric bilinear form, depending on 
a given metric $g$ on $M^{n+1}$ and its derivatives up to order $n+1$. 
The equation
\begin{equation} \label{e1.6}
{\mathcal H}  = 0 
\end{equation}
is conformally invariant, and includes all Einstein metrics (of 
arbitrary signature and $\Lambda$-term). It is a system of $(n+1)^{\rm st}$-order 
equations in the metric, whose leading order term in suitable coordinates 
is of the form $\Box^{\frac{n+1}{2}}$, where $\Box$ is the wave operator of the 
metric $g$. Conformal invariance implies that the system (1.6) is non-degenerate 
at ${\mathcal I}^{+} = \{\Omega  = 0\}$. Theorem 1.1 is then proved by showing 
that natural gauge choices for the diffeomorphism and conformal invariance 
of (1.6) lead again to a symmetrizable system of evolution equations. 

\medskip

 In the context of Theorem 1.2, it is of interest to understand the closure 
$\overline{dS^{\pm}}$ of the space $dS^{\pm}$, i.e. the structure of spacetimes 
which are limits of spacetimes in $dS^{\pm}$ but not themselves in $dS^{\pm}$. 
A first step in this direction was taken in [2] in $3+1$ dimensions, and Theorem 
1.1 allows one to extend this to any even dimension. Let $\overline{dS^{\pm}}$ 
be the closure of $dS^{\pm}$ with respect to the $H^{s+n}\times H^{s}$ topology 
on the Cauchy data on either ${\mathcal I}^{+}$ or ${\mathcal I}^{-}$, i.e. the 
union of the closures with respect to data on ${\mathcal I}^{-}$ and 
${\mathcal I}^{+}$. Let $\partial dS^{\pm} = \overline{dS^{\pm}} \setminus dS^{\pm}$ 
be the resulting boundary consisting of limits of spaces in $dS^{\pm}$ which 
are not in $dS^{\pm}$. 
\begin{theorem} \label{t 1.3.}
  For $(n+1)$ even, a space in the boundary $\partial dS^{\pm}$ of 
$dS^{\pm}$, is described by one of the following three configurations:

 I. A pair of solutions $(M, g^{+}) \in  dS^{+}$ and $(M, g^{-}) \in  
dS^{-}$, each geodesically complete and globally hyperbolic. One has 
${\mathcal I}^{-} = \emptyset$ for $(M, g^{+})$ and ${\mathcal I}^{+} = 
\emptyset$ for $(M, g^{-})$. Both solutions $(M, g^{+})$ and $(M, g^{-})$ 
are ``infinitely far apart''. 

 II. A single geodesically complete and globally hyperbolic solution 
$(M, g) \in  dS^{+}$, either with a partial compactification at ${\mathcal 
I}^{-}$, or ${\mathcal I}^{-} = \emptyset$.

 III. A single geodesically complete and globally hyperbolic solution 
$(M, g) \in  dS^{-}$, either with a partial compactification at ${\mathcal 
I}^{+}$, or ${\mathcal I}^{+} = \emptyset$.
\end{theorem}

  Cases II and III have been distinguished here, but these behaviors become 
identical under a switch of time orientation. 

 One of the main points here is that singularities do not form on 
spaces within $\overline{dS^{\pm}}$. One does expect singularities to 
form ``past'' the boundary $\partial dS^{\pm}$. The most natural limits 
are those of type I; this behavior occurs very clearly and explicitly 
in the family of dS Taub-NUT metrics on ${\mathbb R} \times S^{n}$, 
cf. [2] for further discussion. It would be very interesting to know 
more about the structure of $\overline{dS^{\pm}}$; for instance, is it 
compact and connected?

 Theorems 1.1-1.3 are proved in \S 2, and we close the paper with some 
remarks on extending these results to vacuum equations with $\Lambda  
\leq 0$ and to the Einstein equations coupled to matter fields. 

\medskip

  I would like to thank the referee and Piotr Chru\'sciel for very useful 
comments on the paper. 

\section{Proofs of the Results.}
\setcounter{equation}{0}

 Throughout the paper, we consider globally hyperbolic vacuum 
spacetimes $(M, g)$ with $\Lambda > 0$ in $(n+1)$ dimensions. By 
rescaling if necessary, it is assumed that $\Lambda$ is normalized to 
$\Lambda  = n(n-1)/2$, so that the Einstein equations read
\begin{equation} \label{e2.1}
Ric_{g} = ng. 
\end{equation}
The simplest solution of (2.1) is (pure) deSitter space on $M = {\mathbb R} 
\times S^{n}$, with metric (1.2), or its generalization in (1.3). These de Sitter metrics 
$g_{dS}$ are geodesically complete and globally conformally compact, i.e. in $dS^{\pm}$. 
In fact, defining $s \in (-\frac{\pi}{2}, \frac{\pi}{2})$ 
by $\cosh(t) = \frac{1}{\cos(s)}$ and letting
\begin{equation} \label{e2.2}
\bar g = \cos^{2}(s)g ,
\end{equation}
one has
$$\bar g_{dS} = -ds^{2} + g_{S^{n}(1)},$$
which is the metric on the Einstein static spacetime in the region 
$s  \in  [-\frac{\pi}{2}, \frac{\pi}{2}]$. The metric 
$\bar g_{dS}$ is real analytic on the closure $\bar M = M \cup {\mathcal I}^{+} 
\cup {\mathcal I}^{-}$ and the loci ${\mathcal I}^{+} = \{s  = \frac{\pi}{2}\} = 
\{t = \infty\}, {\mathcal I}^{-} = \{s  = -\frac{\pi}{2}\} = \{t = -\infty\}$ 
represent future and past conformal infinity. The induced metric on 
${\mathcal I}^{\pm}$ is of course the unit round metric $g_{S^{n}(1)}$ on $S^{n}$. 
The same discussion holds for $(N, g_{N})$ as in (1.3) in place of $g_{S^{n}(1)}$. 

\medskip

 Consider Einstein metrics $(M^{n+1}, g)$ in $dS^{+}$, so that there 
is a compactification 
\begin{equation} \label{e2.3}
\bar g = \rho^{2}g
\end{equation}
as in (1.4) to future conformal infinity ${\mathcal I}^{+}$, with 
${\mathcal I}^{+} = \{\rho = 0\}$; all of the analysis below 
works equally well for spaces in $dS^{-}$.

 A compactification $\bar g = \rho^{2}g$ as in (2.3) is called 
geodesic if $\rho (x) = dist_{\bar g}(x, {\mathcal I}^{+})$. These are 
often the simplest compactifications to work with for computational 
purposes. Each choice of boundary metric $\gamma  \in  [\gamma]$ on 
${\mathcal I}^{+}$ determines a unique geodesic defining function $\rho$, 
(and vice versa). The Gauss Lemma gives the splitting
\begin{equation} \label{e2.4}
\bar g = -d\rho^{2} + g_{\rho}, \ \  g = \rho^{-2}(-d\rho^{2} + 
g_{\rho}), 
\end{equation}
where $g_{\rho}$ is a curve of metrics on ${\mathcal I}^{+}$. The 
asymptotic behavior of $g$ at ${\mathcal I}^{+}$ is thus determined by the 
behavior of $g_{\rho}$ as $\rho  \rightarrow 0$. For example, the 
geodesic compactification of the de Sitter metric (1.2) with respect to 
the unit round metric at ${\mathcal I}^{+}$ is
$$\bar g_{dS} = -d\rho^{2} + 
(1+(\frac{\rho}{2})^{2})^{2}g_{S^{n}(1)}, $$
for $\rho  \in  [0,\infty )$. 

 Now consider a Taylor series type expansion for the curve $g_{\rho}$ 
on ${\mathcal I}^{+}$. This was analysed in case of asymptotically 
hyperbolic or AdS metrics with $\Lambda < 0$ by Fefferman-Graham [6], 
and for dS metrics by Starobinsky [19] when $n = 3$. This idea of 
course has further antecedents in the Bondi-Sachs expansion and peeling 
properties of the Weyl tensor when $\Lambda  = 0$. In any case, the FG 
expansion holds equally well for metrics in $dS^{+}$ (or $dS^{-})$ in place 
of asymptotically AdS metrics; in fact the two expansions are very 
closely related, cf. [2], [18] and further references therein. 

 The exact form of the expansion depends on whether $n$ is odd or even. 
If $n$ is odd, then 
\begin{equation} \label{e2.5}
g_{\rho} \sim  g_{(0)} + \rho^{2}g_{(2)} + .... + \rho^{n-1}g_{(n-1)} + 
\rho^{n}g_{(n)} + \rho^{n+1}g_{(n+1)} + ..., 
\end{equation}
with $g_{(0)} = \gamma$. 

 This expansion is even in powers of $\rho$ up to order $n-1$. The 
coefficients $g_{(2k)}$, $0 < k < n/2$ are locally determined by the 
boundary metric $\gamma  = g_{(0)}$; they are explicitly computable 
expressions in the curvature of $\gamma$ and its covariant 
derivatives. For example for $n \geq  3$, 
\begin{equation} \label{e2.6}
g_{(2)} = \frac{1}{n-2}(Ric_{\gamma} - \frac{R_{\gamma}}{2(n-1)}\gamma), 
\end{equation}
cf. also [5], [2] for formulas for $g_{(k)}$ for $k > 2$.

 The term $g_{(n)}$ is transverse-traceless, i.e.
\begin{equation} \label{e2.7}
tr_{\gamma}g_{(n)} = 0,  \ \ \delta_{\gamma}g_{(n)} = 0, 
\end{equation}
but is otherwise undetermined by $\gamma$ and the Einstein equations 
(2.1); thus, at least formally, it is freely specifiable. For $k > n$, 
terms $g_{(k)}$ occur for $k$ both even and odd; the term $g_{(k)}$ 
depends on two boundary derivatives of $g_{(k-2)}$. The main point is 
that all coefficients $g_{(k)}$ are locally computable expressions in 
$g_{(0)}$ and $g_{(n)}$.

 Mathematically, the expansion (2.5) is formal, obtained by 
compactifiying the Einstein equations and taking iterated Lie 
derivatives of $\bar g$ at $\rho = 0$. If the geodesic 
compactification $\bar g$ is in $C^{m,\alpha}(\bar M)$, then 
the expansion holds up to order $m+\alpha$, in the sense that
\begin{equation} \label{e2.8}
g_{\rho} = g_{(0)} + \rho^{2}g_{(2)} + .... + \rho^{m}g_{(m)} + 
O(\rho^{m+\alpha}). 
\end{equation}

 Suppose instead $n$ is even. Then the expansion reads
\begin{equation} \label{e2.9}
g_{\rho} \sim  g_{(0)} + \rho^{2}g_{(2)} + .... + \rho^{n-2}g_{(n-2)} + 
\rho^{n}g_{(n)} + \rho^{n}(\log \rho){\mathcal H}  + ... 
\end{equation}
Again the terms $g_{(2k)}$ up to order $n-2$ are explicitly computable 
from the boundary metric $\gamma$, as is the coefficient ${\mathcal H}$ 
of the $\rho^{n}(\log \rho)$ term. The term $g_{(n)}$ satisfies 
$$tr_{\gamma}g_{(n)} = a, \ \  \delta_{\gamma}g_{(n)} = b,$$
where $a$ and $b$ are explicitly determined by the boundary 
metric $\gamma$ and its derivatives, but $g_{(n)}$ is otherwise 
undetermined by $\gamma$ and the Einstein equations; as before, it is 
formally freely specifiable. The series (2.9) is even in powers of 
$\rho$, (at all orders) and terms of the form $\rho^{2k}(\log \rho )^{j}$ 
appear at order $> n$. Again the coefficients $g_{(k)}$ and ${\mathcal H}_{(k)}$ 
depend on two derivatives of $g_{(k-2)}$ and ${\mathcal H}_{(k-2)}$. 

  Although the expressions (2.5) and (2.9) are only formal in general, 
Fefferman-Graham [6] showed that if the undetermined terms $(g_{(0)}, g_{(n)})$ 
are analytic on the boundary ${\mathcal I}^{+}$, then the expansion (2.5) 
converges, (for $n$ odd), cf. also [2]. Thus $g_{\rho}$ is analytic in $\rho$ for 
$\rho$ small and one has a dS Einstein metric in this region given by (2.4). A 
similar result has recently been proved by Kichenassamy [13], (cf. also [17]), 
for $n$ even; in this case the polyhomogeneous expansion (2.9) converges to $g_{\rho}$ 
for $\rho$ small.

\medskip

 The term ${\mathcal H}$, which appears only when $n$ is even, has a 
number of important interpretations. First Fefferman-Graham [6] observed 
that this tensor, locally computable in terms of the boundary metric 
$\gamma$, is a conformal invariant of $\gamma$ and is (by definition) 
an obstruction to the existence of a formal power series expansion of 
the compactified Einstein metric; in fact it is the only obstruction. 
The tensor ${\mathcal H}$ is also important in the (A)dS/CFT correspondence, 
in that (up a constant) it equals the stress-energy tensor (i.e. the 
metric variation) of the conformal anomaly of the corresponding CFT, 
cf. [5]. It also arises as the stress-energy or metric variation of the 
$Q$-curvature of the boundary metric $\gamma$, cf. [7]. 

 The tensor ${\mathcal H}$ is transverse-traceless 
\begin{equation} \label{e2.10}
\delta{\mathcal H}  = tr {\mathcal H}  = 0, 
\end{equation}
and a conformal invariant of weight $2-n$, i.e. if $\widetilde g = 
\lambda^{2}g$, then $\widetilde {\mathcal H} = \lambda^{2-n}{\mathcal H}$. 
Further, if $g$ is conformal to an Einstein metric, with any value of 
$\Lambda$, then 
\begin{equation} \label{e2.11}
{\mathcal H}  = 0. 
\end{equation}
In addition, as observed in [6], these properties hold for metrics of 
any signature and so the equation (2.11) can be viewed as a conformally 
invariant version of the Einstein equations with an arbitrary $\Lambda$ 
term and arbitrary signature. We are not aware of any analogue of such a 
tensor in odd dimensions. 

\medskip

 We will use the tensor ${\mathcal H}$ to study de Sitter type solutions 
of the Einstein equations (2.1). Although the derivation of the 
obstruction tensor ${\mathcal H}$ arises from the structure at infinity of 
conformally compactified odd-dimensional Einstein metrics, once it is given, 
one can use it to study the Einstein equations themselves in even dimensions. 
Thus, replacing $n$ by $n+1$, a vacuum solution of the Einstein equations 
$(M^{n+1}, g)$ with $\Lambda > 0$, (or any $\Lambda$), in even dimensions is 
a solution of (2.11). For $(M^{n+1}, g) \in dS^{+}$, the equation (2.11), 
being conformally invariant, also holds for the compactified Einstein metric 
$\bar g$ in (2.2); moreover it has the important advantage of being a 
non-degenerate system of equations in $\bar g$. As is well-known [8], 
the translation of the Einstein equations for $(M, g)$ to the compactified 
setting $\bar g$ leads to a degenerate system of equations for $\bar g$.

 When $n = 3$, so $dim M = 4$, up to a constant factor ${\mathcal H}$ is 
the Bach tensor $B$, given by 
\begin{equation} \label{e2.12}
B = D^{*}D(Ric - \frac{R}{6}g) + D^{2}(tr(Ric - \frac{R}{6}g)) + 
{\mathcal R},
\end{equation}
where ${\mathcal R}$ is a term quadratic in the full curvature of $g$. 
(The specific form of ${\mathcal R}$ will not be of concern here). In general, 
for $n \geq 3$ odd, one has, again up to a constant factor, 
\begin{equation} \label{e2.13}
{\mathcal H}  = (D^{*}D)^{\frac{n+1}{2}-2}[D^{*}D(P) + D^{2}(trP)] + 
L(D^{n}\gamma ), 
\end{equation}
where
\begin{equation} \label{e2.14}
P = P(\gamma) = Ric_{g} - \frac{R_{g}}{2n}g,
\end{equation}
cf. [7] for example. This is a system of PDE's in the metric $g$, of order 
$n+1$; $L(D^{n}g)$ denotes lower order terms involving the metric up to 
order $n$. 

 The Bach equation $B = 0$ was originally developed by Bach as a conformally 
invariant version of the Einstein equations (with $\Lambda = 0$), 
and has been extensively studied in this context, cf. [14], [15], [16] for some 
recent work and references therein. It was also used in [3] to study 
regularity properties of conformally compact Riemannian Einstein 
metrics. 

\medskip

 While Einstein metrics, (of any signature and $\Lambda$), are 
solutions of (2.11), of course not all solutions of (2.11) are 
Einstein. In addition, for Lorentzian metrics, ${\mathcal H}$ is not a 
hyperbolic system of PDE's in any of the usual senses; the equation 
(2.11) is invariant under diffeomorphisms and conformal changes of the 
metric, and so requires at least a choice of diffeomorphism and conformal 
gauge to obtain a hyperbolic system. 

 To describe these gauge choices, suppose $(M, g) \in  dS^{+}$, so that $g$ 
is an Einstein metric, satisfying (2.1), and so (2.11), which is asymptotically 
dS to the future. Assume that $(M, g)$ has a geodesic compactification which is 
at least $C^{n}$; then
\begin{equation} \label{e2.15}
\bar g = \rho^{2}g = -d\rho^{2} + g_{\rho}, 
\end{equation}
and $g_{\rho}$ has the expansion (2.8), with $m = n$, $\alpha = 0$. One has 
${\mathcal I}^{+} = \{\rho  = 0\}$ and we set $\gamma  = g_{(0)}$. By the 
solution to the Yamabe problem, one may assume without loss of generality that 
the representative $\gamma \in [\gamma]$ has constant scalar curvature, i.e. 
\begin{equation} \label{e2.16}
R_{\gamma} = const, 
\end{equation}
on ${\mathcal I}^{+}$. However, closer study shows that the operator $P$ in 
(2.14) is not well-behaved in the coordinates adapted to (2.15), i.e. the natural 
geodesic coordinates $(\rho , y_{i})$, where $y_{i}$ are local coordinates on 
${\mathcal I}^{+}$ extended to coordinate functions on $M$ to be invariant under 
the flow of $\nabla\rho$. Further, with this choice of conformal gauge, it is 
difficult to control the scalar curvature $\bar R$ of $\bar g$.

 It is simpest and most natural to choose a conformal gauge of constant 
scalar curvature, (although other choices are possible). Thus, set
\begin{equation} \label{e2.17}
\widetilde g = \sigma^{2}\bar g, 
\end{equation}
where $\sigma$ is chosen to make $\widetilde R = const$. In this gauge, 
the equation (2.13) for $\widetilde g$ simplifies to 
\begin{equation} \label{e2.18}
{\mathcal H}  = (\widetilde {D^{*}D})^{\frac{n+1}{2}-1}\widetilde Ric 
+ L(D^{n}\widetilde g) = 0. 
\end{equation}
The choice of constant for $\widetilde R$ is not important, but it 
simplifies matters if one chooses
\begin{equation} \label{e2.19}
\widetilde R = \bar R|_{{\mathcal I}^{+}} = 
-\frac{n(n-2)}{n-1}R_{\gamma} \equiv c_{0}. 
\end{equation}
The middle equality follows by taking the trace of (2.6), and combining 
this with the Raychaudhuri equation on $\bar g$ and (2.28) below.

 For the diffeomorphism gauge, we choose, as usual, harmonic 
coordinates $x_{\alpha}$ with respect to $\widetilde g$; 
$$\widetilde{\Box} x_{\alpha} = 0. $$
It is assumed that the Cauchy data for $x_{0}$ are such that $x_{0}$ is 
a defining function for ${\mathcal I}^{+}$ near ${\mathcal I}^{+}$, and we 
relabel $x_{0} = t$ so that the coordinates are $(t, x_{i})$, $i = 1, 
..., n$. As usual, Greek letters are used for spacetime indices, while 
Latin is used for spatial indices. Equivalently, but from a slightly 
different point of view, given arbitrary local coordinates 
$x_{\alpha}$, with $x_{0}$ a defining function for the boundary, the 
condition that $x_{\alpha}$ is harmonic with respect to $\widetilde g$ is 
\begin{equation} \label{e2.20}
\widetilde{\Box} x_{\alpha} = 
\partial_{\alpha}\widetilde g^{\alpha\beta}+ 
\frac{1}{2}\widetilde g^{\alpha\beta}\widetilde g^{\mu\nu}\partial_{\alpha}
\widetilde g_{\mu\nu} = 0. 
\end{equation}
In the coordinates $x_{\alpha}$, the metric $\bar g$ in (2.15) 
becomes
\begin{equation} \label{e2.21}
\bar g = g_{00}(dt^{2}) + 2g_{0i}dtdx_{i} + g_{ij}dx_{i}dx_{j}, 
\end{equation}
where 
\begin{equation} \label{e2.22}
g_{00} = -(\partial_{t}\rho )^{2}, \ \   g_{0i} = 
\partial_{t}\rho\partial_{i}\rho ,   \ \ g_{ij} = 
\partial_{i}\rho\partial_{j}\rho  + (g_{\rho})_{ij}. 
\end{equation}
Similarly, for $\widetilde g$, one has
\begin{equation} \label{e2.23}
\widetilde g = \widetilde g_{00}(dt^{2}) + 2\widetilde g_{0i}dtdx_{i} + 
\widetilde g_{ij}dx_{i}dx_{j}, 
\end{equation}
with $\widetilde g_{\alpha\beta} = \sigma^{2}g_{\alpha\beta}.$

 As long as the coordinates are $\widetilde g$-harmonic, the Ricci 
curvature has the form
$$\widetilde Ric_{\alpha\beta} = 
-\frac{1}{2}\widetilde g^{\mu\nu}\partial_{\mu}\partial_{\nu}\widetilde g_
{\alpha\beta} + Q_{\alpha\beta}(\widetilde g, \partial\widetilde g), $$
Similarly at leading order, the Laplacian $\widetilde{D^{*}D}$ has the form 
$\widetilde g^{\mu\nu}\partial_{\mu}\partial_{\nu}$ in harmonic 
coordinates. Thus, with these choices of gauge for the conformal and 
diffeomorphism invariance, the equation $\widetilde {{\mathcal H}} = 0$ has 
the rather simple form
\begin{equation} \label{e2.24}
(\widetilde g^{\mu\nu}\partial_{\mu}\partial_{\nu})^{\frac{n+1}{2}}
\widetilde g_{\alpha\beta} + L(D^{n}\widetilde g) = 0. 
\end{equation}
This is an $N\times N$ system of PDE's for $\widetilde g_{\alpha\beta}$ which is 
diagonal, i.e. uncoupled, at leading order, $N = (n+1)(n+2)/2$. These choices 
for the conformal and diffeomorphism gauges are the simplest; however, they 
are not necessary and other choices, for instance gauges determined by fixed 
gauge source functions, cf. [12], could also be used. 

\medskip

 Having discussed the equations for the metric, we have left to 
determine the equations for $\rho$ in (2.15) and $\sigma$ in (2.17). 
The fact that $\rho$ is a geodesic defining function for 
$\bar g$, i.e. $|\bar \nabla \rho|_{\bar g}^{2} = -1$, implies that 
$$\partial_{t}(g^{\alpha\beta}\partial_{\alpha}\rho\partial_{\beta}\rho) 
= 0, $$
or equivalently,
\begin{equation} \label{e2.25}
\partial_{t}(\sigma^{2}\widetilde g^{\alpha\beta}\partial_{\alpha}\rho
\partial_{\beta}\rho) = 0. 
\end{equation}
To derive the equation for $\sigma$, the equation for the Ricci 
curvature relating $\widetilde g$ and $\bar g$ is
\begin{equation} \label{e2.26}
\bar Ric = \widetilde Ric + 
(n-1)\frac{\widetilde D^{2}\sigma}{\sigma} + 
\{\frac{\widetilde{\Box}\sigma}{\sigma} - n|d\log \sigma|^{2}\}\widetilde g. 
\end{equation}
Taking the trace gives the equation relating the scalar curvatures as
\begin{equation} \label{e2.27}
\sigma^{-2}\bar R = \widetilde R + 
2n\frac{\widetilde{\Box} \sigma}{\sigma} - n(n+1)|d\log \sigma|^{2}_{\widetilde g}. 
\end{equation}
Using the formula analogous to (2.26) relating the Ricci curvature of 
$g$ and $\bar g$, together with the fact that $g$ satisfies (2.1) 
and $\rho$ is a geodesic defining function gives 
\begin{equation} \label{e2.28}
\bar R = -2n\frac{\bar{\Box} \rho}{\rho} = -2n\bar Ric(T,T), 
\end{equation}
where $T = \partial_{\rho} = -\nabla_{\bar g}\rho$. (Observe that the middle 
term in (2.28) is degenerate at ${\mathcal I}^{+}$, since $\rho = 0$ there; 
however, the last term in (2.28) is non-degenerate at $\rho = 0$). Substituting 
(2.28) in (2.27) and using (2.26) gives then the equation 
$$\widetilde R + n(n-1)|d\log \sigma|^{2}_{\widetilde g} = 
-2n\sigma^{-2}[\widetilde Ric(T,T) + 
(n-1)\frac{\widetilde D^{2}\sigma}{\sigma}(T,T)], $$
or equivalently,
\begin{equation} \label{e2.29}
TT(\sigma ) - \langle \widetilde \nabla \sigma , \widetilde \nabla_{T}T \rangle 
= -\frac{1}{n-1}\sigma \widetilde Ric(T,T) - 
\frac{\sigma^{3}}{2n(n-1)}c_{0} - 
\frac{1}{2}\sigma|d\log \sigma|^{2}_{\widetilde g} , 
\end{equation}
where we have also used (2.19).

\medskip

 The equations (2.24), (2.25), and (2.29) represent a coupled system of 
evolution equations for the variables $(\widetilde g_{\alpha\beta}, \rho 
, \sigma)$ on a domain $U$ in $({\mathbb R}^{n+1})^{+}$ with coordinates 
$(t, x_{i})$; the boundary $\partial_{0} U = U \cap\{t=0\}$ corresponds 
to a portion of ${\mathcal I}^{+}$. Written out in more detail, these are: 
\begin{equation} \label{e2.30}
(\widetilde g^{\mu\nu}\partial_{\mu}\partial_{\nu})^{\frac{n+1}{2}}
\widetilde g_{\alpha\beta} = L_{1}(D^{n}\widetilde g)_{\alpha\beta}, 
\end{equation}
\begin{equation} \label{e2.31}
\widetilde g^{00}\partial_{t}\partial_{t}\rho  + 2(\widetilde \nabla^{i}\rho)
(\partial_{i}\partial_{t}\rho ) = L_{2}(D\rho , D\sigma , D\widetilde g), 
\end{equation}
\begin{equation} \label{e2.32}
(\widetilde \nabla^{0}\rho )^{2}\partial_{t}\partial_{t}\sigma  + 
2(\widetilde \nabla^{0}\rho)(\widetilde \nabla^{i}\rho)
\partial_{i}\partial_{t}\sigma  + (\widetilde \nabla^{i}\rho)
(\widetilde \nabla^{j}\rho)\partial_{i}\partial_{j}\sigma  = L_{3}(D\sigma , 
D^{2}\widetilde g). 
\end{equation}
Here $D^{k}w$ denotes derivatives up to order $k$ in the variable $w$ 
and $\widetilde \nabla^{\alpha}\rho $ denotes the $\alpha$-component of 
$\widetilde \nabla\rho$, $\widetilde \nabla\rho  = 
\widetilde g^{\alpha\beta}\partial_{\beta}\rho\partial_{\alpha}$. The terms 
$L_{i}$ are lower order terms. Observe that the system (2.30) for the metric 
$\widetilde g_{\alpha\beta}$ is a closed sub-system, i.e. it does not involve 
$\rho$ or $\sigma$. Moreover, although the equations (2.31) and (2.32) for 
$\rho$ and $\sigma$ are coupled to each other and to (2.30), the system 
(2.30)-(2.32) is uncoupled at leading order. 

\medskip

 Following common practice, we now reduce the system (2.30)-(2.32) to a 
system of $1^{\rm st}$ order equations. There is not a unique way to do 
this, but we will discuss perhaps the simplest method, which uses 
pseudodifferential operators. As usual, the domain $\partial_{0}U \subset 
{\mathbb R}^{n}$ is viewed as a domain in the $n$-torus $T^{n}$ and the 
variables $(\widetilde g_{\alpha\beta}, \sigma, \rho)$ are extended to 
functions on $I \times T^{n}$. 

 Recall that a system of $1^{\rm st}$ order evolution equations 
\begin{equation} \label{e2.33}
\partial_{t}u = \sum_{j=1}^{m} B_{j}(t, x, u)\partial_{j}u + 
c(t,x,u) 
\end{equation}
is symmetrizable in the sense of Lax, cf. [20], [21], if there is a smooth 
matrix valued function $R(t,u,x,\xi)$ on ${\mathbb R} \times {\mathbb R}^{p}
\times T^{*}(T^{n})\setminus 0$, homogeneous of degree 0 in $\xi$, 
such that $R$ is a positive definite $p\times p$ matrix with $R(t,u,x,\xi)\sum 
B_{j}(t,u,x)\xi_{j}$ self-adjoint, for each $(t,u,x,\xi)$. It is 
well-known [20], [21] that strictly hyperbolic systems of PDE, diagonal at 
leading order, are symmetrizable. A symmetrizer $R$ is given by $R = 
\sum P_{k}P_{k}^{*},$ where $P_{k}$ is the projection onto the $k^{\rm th}$ 
eigenspace of the symbol $\sum B_{j}(t,u,x)\xi_{j}$, $1 \leq  k \leq p$.
\begin{proposition} \label{p 2.1.}
  There is a reduction of the system (2.30)-(2.32) to a symmetrizable 
system of $1^{\rm st}$ order evolution equations on $I \times T^{n}$.
\end{proposition}
{\bf Proof:}
 Consider first the closed system (2.30) for $\widetilde g$. This system 
is not strictly hyperbolic; the leading order symbol is diagonal and has 
two distinct real eigenvalues, each of multiplicity $(n+1)/2$. However, the 
eigenspaces of the symbol of $\widetilde{\Box}^{\frac{n+1}{2}}$ vary 
smoothly and do not coalesce. Thus the operator $\widetilde{\Box}^{\frac{n_1}{2}}$ 
is strongly hyperbolic, cf. [12] and references therein. In these circumstances, 
it is essentially standard that the operator $\widetilde{\Box^{k}}$ is symmetrizable, 
for any $k$; for completeness we sketch the proof following [20, \S 5.3]. 

 Let $\widetilde u = \widetilde g_{\alpha\beta}$ be the variable in 
${\mathbb R}^{N}$, $N = (n+1)(n+2)/2$. Write $\Box$, (we drop the tilde here 
and below), in the form 
$$(g^{00})^{-1}\Box  = \partial_{t}^{2} - \sum_{j=0}^{1}A_{j}
(\widetilde u,D_{x})\partial_{t}^{j}, $$
where $A_{j}$ is a differential operator in $x$, homogeneous of order 
$2-j$, depending smoothly $\widetilde u$. Then
\begin{equation} \label{e2.34}
[(g^{00})^{-1}\Box ]^{(n+1)/2} = \partial_{t}^{n+1} - 
\sum_{j=0}^{n}B_{j}(\widetilde u,D_{x})\partial_{t}^{j},
\end{equation}
where $B_{j}$ are differential operators in $x$, homogeneous of order 
$n+1-j$. Set $u_{j} = \partial_{t}^{j}\Lambda^{n-j}\widetilde u$, for $j 
= 0, ..., n$, where $\Lambda  = (1-\Delta)^{1/2}$ and $\Delta$ is the 
standard Laplacian on $T^{n}$. Then (2.30) becomes
$$\partial_{t}u_{j} = \Lambda u_{j+1}, \ 0 \leq  j <  n, \ \ \partial_{t}u_{n} 
= \sum_{j=0}^{n}B_{j}(Pu,D_{x})\Lambda^{j-n}u_{j} + C(Pu), $$
where $Pu = D^{n}\widetilde u$ involves $\widetilde u$ up to $n$ 
derivatives. More precisely, for $\beta +j \leq n$, 
$\partial_{x}^{\beta}\partial_{t}^{j}\widetilde u = 
\partial_{x}^{\beta}\Lambda^{j-n}u_{j};$ for example $\Lambda^{-n}u_{0} 
= \widetilde u$. This is a system of $1^{\rm st}$ order 
pseudodifferential equations in the variables $u = \{u_{j}\}$, $j = 0, 
..., n$ of the form
\begin{equation} \label{e2.35}
\partial_{t}u = L(Pu, D_{x})u + C(Pu).  
\end{equation}
The eigenvalues $\lambda_{\nu}(w, \xi)$ of the matrix $L(w, \xi)$ are the 
roots of the characteristic equation $\tau^{n+1} - \sum B_{j}(w, \xi)\tau^{j}$, 
(up to an overall factor of $i$). Hence, from (2.34), one sees that for each 
$(w, \xi)$, $\xi \neq 0$, there are two distinct roots, each of multiplicity 
$(n+1)/2$. The eigenvalues vary smoothly with $(w, \xi)$ and remain a bounded 
distance apart on the sphere $|\xi| = 1$. The same is true of the corresponding 
eigenspaces. Hence, the system (2.35) has a symmetrizer $R$ constructed in the 
same way as following (2.33), cf. [20, Prop.5.2.C] or [21, Prop.16.2.2]. 

\medskip

 Next we show that the equation (2.31) is also symmetrizable. Let 
$\phi_{i} = -2\widetilde \nabla^{i}\rho /\widetilde g^{00}$. Introducing the 
vector variable $v = (\rho , \rho_{0}, ..., \rho_{n})$ with $\rho_{0} 
= \partial_{t}\rho , \rho_{j} = \partial_{j}\rho$, the equation is 
equivalent to the system
$$\partial_{t}\rho  = \rho_{0}, \ \ \partial_{t}\rho_{0} = 
\sum\phi_{j}\partial_{j}\rho_{0} + c(t,x, v), \ \ \partial_{t}\rho_{j} = 
\partial_{j}\rho_{0}. $$
This has the form
\begin{equation} \label{e2.36}
\partial_{t}v = \sum_{j=1}^{n}B_{j}(x,t,v)\partial_{j}v + 
c(x,t,v), 
\end{equation}
where $B_{j}$ is an $(n+2)\times (n+2)$ matrix with $\phi_{j}$ in the 
$(2,2)$ slot, $1$ in the $(j+2,2)$ slot, and $0$ elsewhere. The system 
(2.36) is coupled at lower order to the equations (2.30) and (2.32) for 
$\widetilde g$ and $\sigma$ respectively, in that $B_{j}$ depends on 
$\widetilde g$ to order 0, while $c$ depends on $\widetilde g$ and 
$\sigma$ to order 1; for the moment, these dependencies are placed in 
the $(x,t)$ dependence of $B_{j}$ and $c$.

 The matrix $\sum B_{j}\xi_{j}$ has the entry $\sum\phi_{j}\xi_{j}$ in 
the $(2,2)$ slot, $\xi_{j}$ in the $(j+2,2)$ slot for $3 \leq j \leq n$, 
and $0$ elsewhere. By a direct but uninteresting computation, it is 
straightforward to see that this matrix is symmetrizable in the sense 
following (2.33).

 Essentially the same argument shows that the equation (2.32) for 
$\sigma$ is again symmetrizable. Thus let $w = (\sigma , \sigma_{0}, 
..., \sigma_{n})$ with $\sigma_{0} = \partial_{t}\sigma$, $\sigma_{j} = 
\partial_{j}\sigma$. The equation (2.32) is equivalent to the system
$$\partial_{t}\sigma  = \sigma_{0}, \ \ \partial_{t}\sigma_{0} = 
\sum\phi_{j}\partial_{j}\sigma_{0} + \sum\psi_{ij}\partial_{j}\sigma_{i} 
+ c(t,x, v), \ \ \partial_{t}\rho_{j} = \partial_{j}\rho_{0}, $$
where $\phi_{j} = -2(\widetilde \nabla^{i}\rho)/|\widetilde \nabla^{0}\rho|$, 
$\psi_{ij} = (\widetilde \nabla^{i}\rho)(\widetilde \nabla^{j}\rho)
/|\widetilde \nabla^{0}\rho|^{2}$. This system has the form
\begin{equation} \label{e2.37}
\partial_{t}w = \sum_{j=1}^{n}B_{j}(x,t,w)\partial_{j}w + 
c(x,t,w), 
\end{equation}
where $B_{j}$ is the $(n+2)\times (n+2)$ matrix with $\phi_{j}$ in the 
$(2,2)$ slot, $1$ in the $(j+2,2)$ slot, and $\psi_{ij}$ in the 
$(i+2,j+2)$ slot. The system (2.37) is again coupled at lower order 
to the equations (2.30) and (2.31) for $\widetilde g$ and $\rho$ 
respectively, in that $c$ depends on $\widetilde g$ to order $2$, 
while $B_{j}$ depends on $\widetilde g$ to order $0$ and $\rho$ 
to order $1$. Again a straightforward but longer (uninteresting) computation 
shows that the matrix $\sum B_{j}\xi_{j}$ is symmetrizable in the sense 
of (2.33). 

 One may then combine the three systems (2.35), (2.36), and (2.37) to a 
single large system in the variable $U = (u, v, w)$. The resulting 
system is then a symmetrizable system of $1^{\rm st}$ order 
pseudodifferential equations, cf. [20], [22].
{\endproof}

 Next consider the Cauchy data for the system (2.30)-(2.32). If one is 
interested in general solutions of this system, then the Cauchy data 
are essentially arbitrary, subject only to the constraint equation 
${\mathcal H}(\bar \nabla \rho, \cdot) = 0$ on ${\mathcal I}^{+}$. 
However, as will be seen in Proposition 2.3 below, it is the specification 
of the Cauchy data which determines the class of conformally Einstein metrics 
among all solutions of (2.30)-(2.32). This is of course closely related to the 
FG expansion (2.8) of the dS metric $g$.

 The Cauchy data for $\sigma$ are
\begin{equation} \label{e2.38}
\sigma  = 1, \ {\rm and} \ \partial_{t}\sigma  = 0 \ \ {\rm at} \ \ {\mathcal I}^{+}, 
\end{equation}
while the Cauchy data for $\rho$ are
\begin{equation} \label{e2.39}
\rho  = 0, \ {\rm and} \ \partial_{t}\rho  = 1 \ \ {\rm at} \ \  {\mathcal I}^{+}. 
\end{equation}

 For the metric $\widetilde g$, the closed subsystem (2.30) is of order 
$n+1$, so Cauchy data are specified by prescribing 
$(\partial_{t})^{k}\widetilde g_{\alpha\beta}$, $k = 0, 1, ..., n$ at 
${\mathcal I}^{+}$. We compute the data inductively. First, the condition 
(2.38) implies that $\widetilde g_{ij} = \bar g_{ij}$ at ${\mathcal I}^{+}$. 
Thus at order 0, set 
\begin{equation} \label{e2.40}
\widetilde g_{00} = -1, \ \ \widetilde g_{0i} = 0, \ \ \widetilde g_{ij} = 
\gamma_{ij} \ \ {\rm at} \ \  {\mathcal I}^{+}, 
\end{equation}
since $\rho_{i} = 0$ at ${\mathcal I}^{+}$.

 At $1^{\rm st}$ order, (2.38) and (2.39) together with (2.23) show that 
$\partial_{t}\widetilde g_{ij} = \partial_{t}\bar g_{ij}$ at ${\mathcal 
I}^{+}$, and the FG expansion (2.8) gives $\partial_{t}\bar g_{ij} 
= 0$ at ${\mathcal I}^{+}$. Thus, set 
\begin{equation} \label{e2.41}
\partial_{t}\widetilde g_{ij} = 0 \ \ {\rm at} \ \  {\mathcal I}^{+}. 
\end{equation}
(This condition, and related ones below, are necessary to obtain 
Einstein metrics). The first derivatives of the mixed components 
$\widetilde g_{0\alpha}$ of $\widetilde g$ are determined by the requirement 
that the coordinates $x_{\alpha} = (t, x_{i})$ are harmonic at ${\mathcal I}^{+}$ 
with respect to $\widetilde g$, i.e. for each $\beta$, 
\begin{equation} \label{e2.42}
\partial_{\alpha}\widetilde g^{\alpha\beta}+ 
\frac{1}{2}\widetilde g^{\alpha\beta}\widetilde g^{\mu\nu}\partial_{\alpha
}\widetilde g_{\mu\nu} = 0. 
\end{equation}
Via (2.40)-(2.41), this determines $\partial_{t}\widetilde g_{0\alpha}$ at 
${\mathcal I}^{+}$.

 At $2^{\rm nd}$ order, the equation (2.27) implies, using the normalization 
(2.19), that $\partial_{t}^{2}\sigma = 0$ at ${\mathcal I}^{+}$. Also, 
$\partial_{t}^{2}(\rho_{i}\rho_{j}) = 0$, and hence, from the FG 
expansion (2.8), we set  
\begin{equation} \label{e2.43}
\partial_{t}^{2}\widetilde g_{ij} = 2g_{(2)} \ \ {\rm at} \ \  {\mathcal I}^{+}. 
\end{equation}
The $2^{\rm nd}$ derivatives $\partial_{t}\widetilde g_{0\alpha}$ are then 
determined by (2.43), the lower order Cauchy data, (2.38)-(2.41), and the 
$t$-derivative of (2.42) at $t = 0$.

 At $3^{\rm rd}$ order, suppose first $n = 3$, so that $dim M = 4$. Then a 
straightforward computation, using (2.8), the Raychaudhuri equation on 
$\bar g$ and (2.28), shows that $\partial_{t}\bar R = 
6ntrg_{(3)} = 0$ at ${\mathcal I}^{+}$, where the last equality follows 
from (2.7). Hence, (2.27) gives $\partial_{t}^{3}\sigma = 0$. 
Similarly, (2.39) gives $\partial_{t}^{3}(\rho_{i}\rho_{j}) = 0$. Thus, 
set 
\begin{equation} \label{e2.44}
\partial_{t}^{3}\widetilde g_{ij} = 6g_{(3)} \ \ {\rm at} \ \  {\mathcal I}^{+}. 
\end{equation}
This term is free or unconstrained, subject to the transverse-traceless 
constraint (2.7). As before the mixed term at order 3, 
$\partial_{t}^{3}\widetilde g_{0\alpha}$ is determined by taking two 
$t$-derivatives of (2.42) at $t = 0$, and using (2.44) together with the 
determination of the lower order Cauchy data. 

 Suppose instead $n > 3$ and hence $n \geq 5$. Then $g_{(3)} = 0$ and 
same arguments as above give 
\begin{equation} \label{e2.45}
\partial_{t}^{3}\widetilde g_{ij} = 0 \ \ {\rm at} \ \  {\mathcal I}^{+}, 
\end{equation}
with $\partial_{t}^{3}\widetilde g_{0\alpha}$ again determined from two 
$t$-derivatives of (2.42), (2.45) and lower order Cauchy data. 

 At $4^{\rm th}$ order on ${\mathcal I}^{+}$, (assuming $n \geq 5$), by (2.27) and 
the fact that $\partial_{t}^{k}\sigma = 0$, $k \leq 3$ on ${\mathcal I}^{+}$, 
one has $\partial_{t}^{4}\sigma  = -2n\partial_{t}^{2}\bar R$, and again 
computations as above then give $\partial_{t}^{2}\bar R = 24ntr g_{(4)}$. 
Also, taking $i$-derivatives of (2.31) or (2.25) and using (2.38)-(2.39) 
shows that $\partial_{t}^{4}(\rho_{i}\rho_{j}) = 0$ at ${\mathcal I}^{+}$. 
It follows that, at $t = 0$,
\begin{equation} \label{e2.46}
\partial_{t}^{4}\widetilde g_{ij} = 24(g_{(4)} - 2n^{2}tr g_{(4)})\gamma . 
\end{equation}
Again, $\partial_{t}^{4}\widetilde g_{0\alpha}$ is determined by taking 
three $t$-derivatives of (2.42) at $t = 0$, and using (2.46) with the 
determination of the lower order Cauchy data.

 At $5^{\rm th}$ order, suppose $n = 5$. As in the case $n = 3$, 
$\partial_{t}^{3}\bar R = ctrg_{(5)} = 0$ at ${\mathcal I}^{+}$ while 
$\partial_{t}^{5}(\rho_{i}\rho_{j}) = 0$. Hence, as in the case $n =3$, 
\begin{equation} \label{e2.47}
\partial_{t}^{5}\widetilde g_{ij} = (5!)g_{(5)}, 
\end{equation}
which is freely specifiable, subject to the transverse-traceless 
constraint. As before the mixed term at order 5, 
$\partial_{t}^{5}\widetilde g_{0\alpha}$ is determined by taking four 
$t$-derivatives of (2.42) at $t = 0$, and using (2.47) with the 
determination of the lower order Cauchy data. 

 If $n > 5$, then as before,
\begin{equation} \label{e2.48}
\partial_{t}^{5}\widetilde g_{ij} = 0, 
\end{equation}
with $\partial_{t}^{5}\widetilde g_{0\alpha}$ again determined from 
(2.42). It is clear that one can continue inductively in this way to 
determine the Cauchy data $\partial_{t}^{k}\widetilde g_{\alpha\beta}$ 
up to order $n$. Since $\partial_{t}^{6}(\rho_{i}\rho_{j}) \neq 0$ at 
$t = 0$, these and higher derivative terms contribute to the Cauchy data 
at order 6 and above. However, one sees by differentiations of (2.31) and 
(2.42) that these terms are all determined by lower order Cauchy data for 
$\widetilde g$. 

 In sum, Cauchy data for $\widetilde g_{\alpha\beta}$, i.e. the data 
$\partial_{t}^{k}\widetilde g_{\alpha\beta}$, $0 \leq  k \leq n$, are 
determined by the Cauchy data (2.38), (2.39) for $\rho$ and $\sigma$, 
the equations (2.31)-(2.32), (or (2.25), (2.27)) for $\rho$ and 
$\sigma$, the harmonic equation (2.42), and the coefficients $g_{(k)}$ 
in the FG expansion (2.8). Thus, the Cauchy data are uniquely 
determined in terms of the free data 
\begin{equation} \label{e2.49}
\gamma  = g_{(0)}, \ \ {\rm and} \ \  \tau = g_{(n)}, 
\end{equation}
which are arbitrary, subject to the transverse-traceless constraint 
(2.7) on $g_{(n)}$ and the constant scalar curvature constraint (2.16) 
or (2.19) on the representative $\gamma \in [\gamma]$. Abusing notation 
slightly, we will call $(\gamma, \tau)$ Cauchy data for $\widetilde g_{\alpha\beta}$, 
since this data determines the rest of the Cauchy data 
$\partial_{t}^{k}\widetilde g_{\alpha\beta}$, $0 < k < n$ uniquely. 

 The analysis above then gives:
\begin{proposition} \label{p 2.2.}
  The system (2.30)-(2.32) for $(\widetilde g_{\alpha\beta}, \rho ,\sigma)$ 
is well-posed in $H^{s+n}({\mathcal I}^{+})\times H^{s}({\mathcal I}^{+})$, 
$s > \frac{n}{2} + 2$. Thus, given Cauchy data $(\gamma , \tau) \in  
(H^{s+n}({\mathcal I}^{+}), H^{s}({\mathcal I}^{+}))$ 
satisfying (2.38), (2.39) and (2.49), and satisfying the constraints 
(2.7), (2.19), there is a unique solution 
$(\widetilde g_{\alpha\beta}, \rho , \sigma)$ of (2.30)-(2.32) with
\begin{equation} \label{e2.50}
(\widetilde g_{\alpha\beta}, \rho , \sigma ) \in  C(I, H^{s+n}({\mathcal I}^{+})) 
\cap  C^{n}(I, H^{s}({\mathcal I}^{+})).
\end{equation}
Further, if $(\gamma_{s}, \tau_{s})$ is a continuous curve in 
$H^{s+n}({\mathcal I}^{+})\times H^{s}({\mathcal I}^{+})$, then the solutions 
$(\widetilde g_{s}, \rho_{s}, \sigma_{s})$ vary continuously with $s$.
\end{proposition}

{\bf Proof:} In the local coordinates $(t, x_{i})$, the system (2.30)-(2.32) is 
symmetrizable and so for given Cauchy data on $T^{n}$, it has a unique solution 
on $I\times T^{n}$ satisfying (2.50), with $T^{n}$ in place of ${\mathcal I}^{+}$, 
cf. [20, \S 5.2-5.3] or also [21], [22]. The existence of such local solutions holds 
for $s > \frac{n}{2} + 1$. By restriction, one thus obtains local 
solutions on the domain $I \times U \subset I \times T^{n}$, for $U \subset 
{\mathcal I}^{+}$ as preceding (2.30). 

  To prove these local solutions obtained from domains on $I\times {\mathcal I}^{+}$ 
patch together to give a unique solution on $I\times {\mathcal I}^{+}$, it is 
necessary and sufficient to prove that the system (2.30)-(2.32) has finite domains 
of dependence (or equivalently uniqueness in the local Cauchy problem). This is 
well-known to be true for symmetric hyperbolic systems of PDE's, (cf. [22, \S IV.4] 
for example); however, the reduction of (2.30) is to a symmetric system of first order 
pseudodifferential equations, for which finite propagation speed is not true in 
general. Nevertheless, standard methods show that solutions of (2.30)-(2.32) 
do have local uniqueness in the Cauchy problem.

  Thus, consider the closed subsystem (2.30) for $\widetilde g$. A standard 
argument using the mean-value theorem shows that it suffices to prove local 
uniqueness in the Cauchy problem for the associated linear equation $L(\widetilde g) 
= 0$, where the coefficients of (2.30) are frozen at a given metric $\widetilde g_{0}$, 
cf. [12] for instance. The linear operator $L$ is self-adjoint at leading order, 
and hence by Proposition 2.1, one has existence and uniqueness of solutions to the 
Cauchy problem on $I\times T^{n}$ for the adjoint equation $L^{*}u = \phi$. It is 
then  well-known, cf. [22, Thm.IV.4.3], that this suffices to prove local uniqueness 
in the Cauchy problem for $L$, and hence for the system (2.30). Given the local 
uniqueness for $\widetilde g$, exactly the same method can be applied to (2.31)-(2.32) 
to give local uniqueness for $\rho$ and $\sigma$. 

  It follows that the system (2.30)-(2.32) does have finite propagation speed, 
and hence the local solutions patch together uniquely to give a unique global 
solution on $I \times {\mathcal I}^{+}$ satisfying (2.50). As is well-known 
in the case of Einstein metrics, (cf. [12] for example), the patching of local 
coordinate charts  requires an extra derivative; the same analysis holds here, 
so that we assume $s > \frac{n}{2} + 2$. 

  A lower bound for the time of existence $I = [0, t_{0})$ of the solution 
depends only on an upper bound on the norm of the initial data in $H^{s+n} 
\times H^{s}$. This implies that the last statement is an immediate 
consequence of the existence and uniqueness theorem. 

{\endproof}

 Given a solution $(\widetilde g_{\alpha\beta}, \rho , \sigma)$ of the 
Cauchy problem, one may then construct the ``physical'' metric $g$ by setting
\begin{equation} \label{e2.51}
g = \rho^{-2}\sigma^{-2}\widetilde g. 
\end{equation}
Since $\sigma$ is bounded away from 0 and $\infty$ near ${\mathcal I}^{+}$ and 
$\rho$ is a geodesic defining function, it is easy to see that 
the metric $g$ is future geodesically complete, i.e. geodesically 
complete to the future of some Cauchy surface $\Sigma  \subset  
(M^{n+1}, g)$. However, it is not so clear that the metric $g$ 
is Einstein, or equivalently that the metric $\widetilde g$ is conformally 
Einstein. For this, one needs to verify first that the gauge  
condition (2.42) that the coordinates remain harmonic, is 
preserved for the solution $\widetilde g$. If this is so, it then 
follows that $\widetilde g$ is a solution of the equation (2.11). 
Secondly, the equation (2.11) admits solutions which are not 
conformally Einstein, and so one needs to verify that the constructed 
solution $\widetilde g$ actually is conformally Einstein. 

 Both of these conditions on $\widetilde g$ can be verified by 
computation; however, the computations will be somewhat long and involved. 
Instead, we verify both conditions together by using the following 
simple conceptual technique, based on analyticity.

\begin{proposition} \label{p 2.3.}
  Any solution $(\widetilde g_{\alpha\beta}, \rho , \sigma )$ of the 
Cauchy problem in Proposition 2.2 defines an Einstein metric in 
$dS^{+}$ via (2.51).
\end{proposition}

{\bf Proof:} Suppose first that the free data $(\gamma , g_{(n)})$ are 
analytic on ${\mathcal I}^{+}$ (or on a domain in ${\mathcal I}^{+}$). 
As noted above, the expansion (2.5) then converges to $g_{\rho}$ and gives 
a solution, denoted $g_{E}$, to the Einstein equations (1.1); this metric 
has the form (2.4), with compactification $\bar g_{E}$, in a neighborhood of 
${\mathcal I}^{+}$. In particular, both $\bar g_{E}$ and $\rho$ are analytic 
near ${\mathcal I}^{+}$. Moreover, since the coefficients of the equation (2.27) 
defining $\sigma$ are then also analytic, and since the Cauchy data (2.38) for 
$\sigma$ are analytic, the Cauchy-Kowalewsky theorem ([21, \S 16.4]) shows that 
$\sigma$ and hence $\widetilde g_{E} = \sigma^{2}\bar g_{E}$ are analytic near 
${\mathcal I}^{+}$. Of course the metric $\widetilde g_{E}$ is a solution of (2.11), 
and hence a solution of (2.30)-(2.32). with Cauchy data determined by (2.38)-(2.39) 
and $(\gamma, g_{(n)})$. 

  On the other hand, let $(\widetilde g, \rho , \sigma)$ be the solution to the 
Cauchy problem (2.30)-(2.32) with Cauchy data determined by (2.38)-(2.39) and 
$(\gamma, g_{(n)})$ given by Proposition 2.2. Since $(\widetilde g_{E}, \rho , \sigma)$ 
is also a solution of this Cauchy problem, with the same initial data, it follows from 
the uniqueness part of Proposition 2.2 that $\widetilde g_{E} = \widetilde g$ in a 
neighborhood of ${\mathcal I}^{+}$. Moreover, the results in [1] show that 
$\widetilde g_{E}$ remains analytic within its globally hyperbolic development, 
so that $\widetilde g_{E} = \widetilde g$ everywhere in the domain of $\widetilde g$. 
Hence $g = \rho^{-2}\sigma^{-2}\widetilde g$ is an Einstein metric in $dS^{+}$, 
realizing the given data on ${\mathcal I}^{+}$. (In particular, this shows that the 
harmonic gauge (2.42) must be preserved for analytic initial data). 

 Now analytic data $(\gamma , g_{(n)})$ are dense in $H^{s+n}\times H^{s}$ 
data with respect to the $H^{s+n}\times H^{s}$ topology on ${\mathcal I}^{+}$. 
The Cauchy stability given by Proposition 2.2 implies that if 
$(\gamma_{i}, g_{(n),i})$ is a sequence of analytic Cauchy data 
converging in $H^{s+n}\times H^{s}$ to $H^{s+n}\times H^{s}$ data 
$(\gamma , g_{(n)})$, then the corresponding solutions $(\widetilde g_{i}, 
\rho_{i}, \sigma_{i})$ also converge to the (unique) solution 
$(\widetilde g, \rho , \sigma)$ with Cauchy data $(\gamma , g_{(n)})$. 
Hence the metric $g$ in (2.51) is Einstein.

{\endproof}

 Propositions 2.2-2.3 give the existence of an Einstein metric 
$(M^{n+1}, g) \in  dS^{+}$, with arbitrarily prescribed asymptotic 
behavior $(\gamma , \tau)$ on ${\mathcal I}^{+}$, subject to the 
constraints (2.7) and (2.19). Suppose $[\gamma'] = [\gamma]$ and 
$[\tau'] = [\tau]$ as following (1.5), so that $\gamma' = \lambda^{2}\gamma$ 
and $\tau' = f(\lambda)\tau$. Given $\lambda$, there exists a diffeomorphism 
$\phi: \bar M \rightarrow \bar M$, with $\phi|_{{\mathcal I}^{+}} = id$, such that 
$\lim_{\rho \rightarrow 0}\phi^{*}(\rho) / \rho = \lambda^{-1}$, where $\rho$ 
is the geodesic defining function determined by $\gamma$. Setting $g' = 
\phi^{*}g$, one has $\bar g' = \rho^{2}\phi^{*}(g) = \phi^{*}(\bar g)
(\frac{\rho}{\phi^{*}(\rho)})^{2}$. Hence, the boundary metric of $\bar g'$ 
equals $\gamma'$. Similarly, by the uniqueness, any Einstein metric 
$(M^{n+1}, g')$ with Cauchy data $(\gamma', \tau')$ satisfying (1.5) 
differs from $(M^{n+1}, g)$ by a diffeomorphism $\phi$ equal to the 
identity on ${\mathcal I}^{+}$. This completes the proof of Theorem 1.1.

{\endproof}

\begin{remark} \label{r2.4}
{\rm Theorem 1.1 is formulated as a global result, in the sense that 
future conformal infinity ${\mathcal I}^{+}$ is a compact smooth 
manifold. However, Propositions 2.1-2.3 all hold locally, by the 
finite propagation speed of the system (2.30)-(2.32). Hence, 
Theorem 1.1 also holds locally, where ${\mathcal I}^{+}$ is an 
open manifold with a finite number of local charts. Of course the 
uniqueness statement then holds only within the domain of dependence 
of the initial data. 

  Many of the standard solutions of the Einstein equations (2.1) have 
${\mathcal I}$ non-compact; this is the case for instance for the dS 
Schwarzschild metrics. }
\end{remark}

{\bf Proof of Theorem 1.2.}

 Let $(M, g_{0})$ be a dS Einstein metric in $dS^{\pm}$ with Cauchy 
data $([\gamma^{+}], [g_{(n)}^{+}])$ and $([\gamma^{-}], [g_{(n)}^{-}])$ 
induced on ${\mathcal I}^{+}$ and ${\mathcal I}^{-}$ respectively. Thus, 
there exists a smooth defining function $\Omega: \bar M \rightarrow 
{\mathbb R}$ such that $\bar g_{0} = \Omega^{2}g_{0}$ extends to a metric 
on $\bar M = M \cup {\mathcal I}^{+} \cup {\mathcal I}^{-} \simeq 
I \times \Sigma$; here $\Sigma$ is a Cauchy surface for $(M, g_{0})$ and 
$I$ is a {\it compact} time interval. We will choose $\Omega$ to be 
a geodesic defining function $\rho$ in a neighborhood of ${\mathcal I}^{+}$ 
and ${\mathcal I}^{-}$ so that $\bar g_{0}$ is $C(I\times H^{s+n}(\Sigma)) 
\cap C^{n}(I\times H^{s}(\Sigma))$ up to $\bar M$. The choice of $\Omega$ 
defines representatives $(\gamma^{+}, g_{(n)}^{+})$, $(\gamma^{-}, g_{(n)}^{-})$ 
in the conformal classes $([\gamma^{+}], [g_{(n)}^{+}])$, $([\gamma^{-}], 
[g_{(n)}^{-}])$. In the following, we work in the $H^{s+n}\times H^{s}$ topology.

  Let ${\mathcal U}^{+}$ be an open neighborhood of Cauchy data on ${\mathcal I}^{+}$ 
containing the given data $(\gamma^{+}, g_{(n)}^{+})$. Then for all data 
$(\hat \gamma^{+}, \hat g_{(n)}^{+}) \in  {\mathcal U}^{+}$, there exists 
$T < \infty$, depending on ${\mathcal U}^{+}$, such that the maximal globally 
hyperbolic dS Einstein metric $(M^{n+1}, \hat g)$ having Cauchy data 
$(\hat \gamma^{+}, \hat g_{(n)}^{+})$ given by Theorem 1.1 is defined on 
$[T,\infty )\times \Sigma$; here the time factor is proper time $t = 
-\log \frac{\rho}{2}$, where $\rho$ is the geodesic defining function. 
The Cauchy data of such solutions $\hat g$ at $\Sigma  = \{T\}\times \Sigma$ 
then forms an open set ${\mathcal U}^{T}$ in the space of Cauchy data for the 
Einstein equations (2.1) on $\Sigma$. By passing to an open subset 
${\mathcal V}^{T} \subset {\mathcal U}^{T}$ if necessary, the Cauchy 
stability theorem for the (standard) Einstein equations implies that the 
maximal globally hyperbolic development of any $\hat g$ with data in 
${\mathcal V}^{T}$ contains the region $[-T, T]\times \Sigma$, and induces 
again an open set of Cauchy data ${\mathcal V}^{-T}$ at $\{-T\}\times \Sigma$. 
Then, as above with ${\mathcal I}^{+}$, there is an open set ${\mathcal U}^{-}$ 
of Cauchy data on ${\mathcal I}^{-}$ whose future development gives a 
non-empty open subset of ${\mathcal V}^{-T}$. Combining these three 
unique developments gives a global solution $(M^{n+1}, \hat g) \in 
dS^{\pm}$, which completes the proof of Theorem 1.2.

{\endproof}
\begin{remark} \label{r2.5}
{\rm The proof above shows that the space $dS^{\pm}$ is also stable with respect to 
perturbations of the Cauchy data $(\Sigma, \gamma, K)$, (satisfying the constraint 
equations), on a compact Cauchy surface $\Sigma \subset (M^{n+1}, g)$ in the 
$H^{s+n}\times H^{s+n-1}$ topology on $\Sigma$. It is well-known, cf. [4] that 
this is not the case for perturbations of asymptotically flat Cauchy data when 
$\Lambda = 0$, in that smoothness of the resulting space-time at conformal infinity 
is lost for generic perturbations. }
\end{remark}

{\bf Proof of Theorem 1.3.}

 This result is proved for $n = 3$ in [2], and it is pointed out there 
that the same proof holds provided one has the Cauchy stability result of 
Theorem 1.1, (i.e. Friedrich's result [8] when $n = 3$). Given then Theorem 1.1, 
the proof of Theorem 1.3 is exactly the same as that given in [2], to which we 
refer for details. 

{\endproof}

\begin{remark} \label{r 2.6.}
  {\rm This paper has focussed on the de Sitter case $\Lambda > 0$ mainly 
for simplicity, but also because there are no direct analogues of 
Theorems 1.2 or 1.3 when $\Lambda = 0$ or $\Lambda < 0$, due to the more 
complicated nature of conformal infinity. Nevertheless, one expects that the 
analogues of Theorem 1.1 for $\Lambda = 0$ and $\Lambda < 0$, as formulated and 
proved by Friedrich [8] in the case $n = 3$, hold for all even dimensions. When 
$\Lambda < 0$, future space-like infinity ${\mathcal I}^{+}$ is replaced 
by time-like infinity ${\mathcal I}$, while when $\Lambda = 0$, ${\mathcal I}^{+}$ 
is replaced by future null infinity. 

  We hope to discuss these situations elsewhere. } 
\end{remark}

\begin{remark} \label{r 2.7.}
 {\rm  We close with a brief remark on the applicability of the methods used 
above to the Einstein equations coupled to other matter fields. As 
noted above, up to multiplicative constants, the tensor ${\mathcal H} $ is 
the metric variation, or stress-energy tensor, of the conformal anomaly 
or of the $Q$-curvature. The conformal invariance of these functionals 
corresponds to the conformal invariance of ${\mathcal H}$. For functionals 
containing the metric coupled to other fields which are conformally 
invariant, and whose field equations are symmetric hyperbolic, it seems 
very likely that the methods used above will again lead to a well-posed 
Cauchy problem at ${\mathcal I}^{+}$, as in Theorem 1.1. In dimensions 
$3+1$, this is the case for the Einstein equations coupled to gauge 
fields, i.e. Einstein-Maxwell or Einstein-Yang-Mills fields. Theorem 
1.1 has already been proved in this situation by Friedrich [11], and so 
at best one would have a different method of proof of this result. In 
higher dimensions, the EM or YM action is not conformally invariant, 
and it is less clear if the method can be adapted to this situation. }
\end{remark}

\bibliographystyle{plain}

\medskip

\begin{center}
August, 2004/November, 2004
\end{center}

\medskip
\noindent
\address{Department of Mathematics\\
S.U.N.Y. at Stony Brook\\
Stony Brook, NY 11794-3651}

\noindent
\email{anderson@math.sunysb.edu}

\end{document}